\documentclass[aps,pra,twocolumn,superscriptaddress]{revtex4-2}

\usepackage{mathtools}
\usepackage{amssymb}
\usepackage{graphicx}
\usepackage{bbold}
\usepackage{hyperref}
\usepackage{xcolor}
\usepackage{enumerate}

\usepackage{xcolor}

\usepackage{soul}

\hypersetup{colorlinks=true, linkcolor=blue, urlcolor=blue, citecolor=blue}

\newcommand{\sys}{\mathcal S}
\newcommand{\env}{\mathcal E}
\newcommand{\uba}{Departamento de F\'\i sica, FCEyN, UBA, Pabell\'on 1, Ciudad Universitaria, 1428 Buenos Aires, Argentina}
\newcommand{\ifiba}{Instituto de F\'\i sica de Buenos Aires, UBA CONICET, Pabell\'on 1, Ciudad Universitaria, 1428 Buenos Aires, Argentina}

\begin{document}
\title{Time-extensive classical and quantum correlations in thermal machines}

\author{Milton Aguilar}
	\email{mil@df.uba.ar}
	\affiliation{\uba}
	\affiliation{\ifiba}

\author{Juan Pablo Paz}
	\email{paz@df.uba.ar}
	\affiliation{\uba}
	\affiliation{\ifiba}

\date{\today}

\begin{abstract}
We study intraenvironmental classical and quantum correlations in a thermal machine, which is modeled as a driven quantum system coupled with thermal reservoirs. We compute the mutual information, the quantum discord, and the entanglement between two parts of the environment formed by oscillators centered around two different frequencies. We show that there are only two processes that generate time-extensive correlations in the long-time limit. First, there is a resonant process which is responsible for the transport of excitations between different environmental modes due to the absorption (or emission) of energy from (or into) the driving field. Second, there is a nonresonant process that transforms the energy from the external driving into pairs of excitations in two environmental modes. We show that there is a regime when the mutual information and the quantum discord between the parts of the environment correlated by these two processes grow quadratically in time, while entanglement production is time-extensive.
\end{abstract}

\maketitle

\section{Introduction}

The tools and resources of information theory have brought astounding advancements to our understanding of the quantum world, and have permeated all the layers of quantum theory. In recent years, the connection between quantum information \cite{nielsenChuang} and quantum thermodynamics \cite{andersVinjanampathy} has been extensively explored, and in the heart of it all lies the study of correlations between physical systems. This has provided, for example, much necessary insight into the emergence of thermodynamic irreversibility from unitary dynamics \cite{espositoLindenbergBroeck, ptaszynskiEsposito}, and a generalization of the laws of thermodynamics that resolve their apparent violations in correlated scenarios \cite{beraRieraLewenstein} (such as anomalous heat flows from cold to hot baths \cite{jenningsRudolph}, and memory erasure accompanied by work extraction instead of heat dissipation \cite{rioAbergRenner}).

In view of their seemingly fundamental role, it is natural to wonder if correlations can be exploited as a thermodynamic resource. In the context of single-shot thermodynamics \cite{aberg}, both classical and quantum correlations have been studied in their role as a resource in state formation or work extraction \cite{sapienzaCerisolaRoncaglia,muller,perarnaullobetHovhannisyanHuber,francicaGooldPlastina}. In addition, considering that separability has often been used synonymously with classicality, entanglement in particular has been put in the spotlight in order to understand how nonclassical correlations affect thermal machines \cite{aguilarFreitasPaz,brunnerHuberLinden,bohrBraskHaackBrunner,khandelwalPalazzoBrunner}.

Understanding the nature and creation of correlations is a necessary task to further advance our knowledge of the foundations of thermodynamics. In this work we address these issues by studying how and what type of correlations are built up between different parts of the environment of a thermal machine while it is continuously driven beyond the transient regime. We will study a periodically driven system $\sys$ in contact with two bosonic environments, $\env_{R}$ and $\env_{L}$, at different temperatures. First, we show that there are only two physical processes that constantly create intraenvironmental time-extensive correlations in the long-time limit. On the one hand, the driving field can transport an excitation from one mode to another one. On the other, the energy of the driving can split and create a pair of excitations that are dumped in different modes. The first process is none other than the quantum manifestation of the heat current from classical thermodynamics, that flows either from the hotter reservoir to the colder one in a heat engine or the reverse in a heat pump. The second one is at the core of the third law of thermodynamics: it always heats up the environment, preventing it from reaching absolute zero temperature. As a consequence of these two processes, time extensive classical and quantum correlations are established only between parts of the environment centered around frequencies such that their sum or difference is  a multiple of the driving frequency $\omega_{d}$. And second, we analyze the nature of those correlations. We present simple, analytical expressions for the quantification of classical and quantum correlations and analyze their relation and their interplay. We also show that the ``classical'' transport of excitations swaps the entanglement created by the creation of a pair of excitations to other modes.

The study of intraenvironmental correlations is not plentiful in the literature of quantum thermodynamics. Recent findings \cite{espositoLindenbergBroeck, ptaszynskiEsposito} show that thermodynamic irreversibility is intrinsically linked to the development of correlations in the environment. Indeed, in Refs. \cite{espositoLindenbergBroeck, ptaszynskiEsposito} it was shown that these correlations are responsible for the entropy production in the long-time limit, using a fermionic system as an example. Our work can also be interpreted as providing support for this idea. Thus, we obtain similar results in a bosonic model, which is the paradigmatic model of a thermal machine. This paper also contains a generalization of previous results reported in Ref. \cite{aguilarFreitasPaz} where the generation of entanglement in the environment was studied. Here we study other measures of classical and quantum correlations, such as the mutual information and the quantum discord. We also present novel results and interpretations about the nature and behavior of the entanglement generated in thermal machines.

The paper is organized as follows: In Sec. II we give a brief overview of the model and its solution. It is a generalization of the standard quantum Brownian motion model including a time-dependant driving field. In Sec. III we use the previous results to compute the covariance matrix between different environmental modes and we discuss its properties in the long-time regime. In Sec. IV we present the measures we will use later to quantify and describe classical and quantum correlations, including entanglement. In Sec. V we study classical and quantum correlations between environmental bands at an arbitrary environmental temperature. We provide simple analytical expressions to compute the mutual information, the quantum discord and the entanglement in terms of physically meaningful parameters. We summarize our results in Sec. VI.

\section{The model}
\begin{figure}[htp]
    \begin{center}
    \includegraphics[scale=.5]{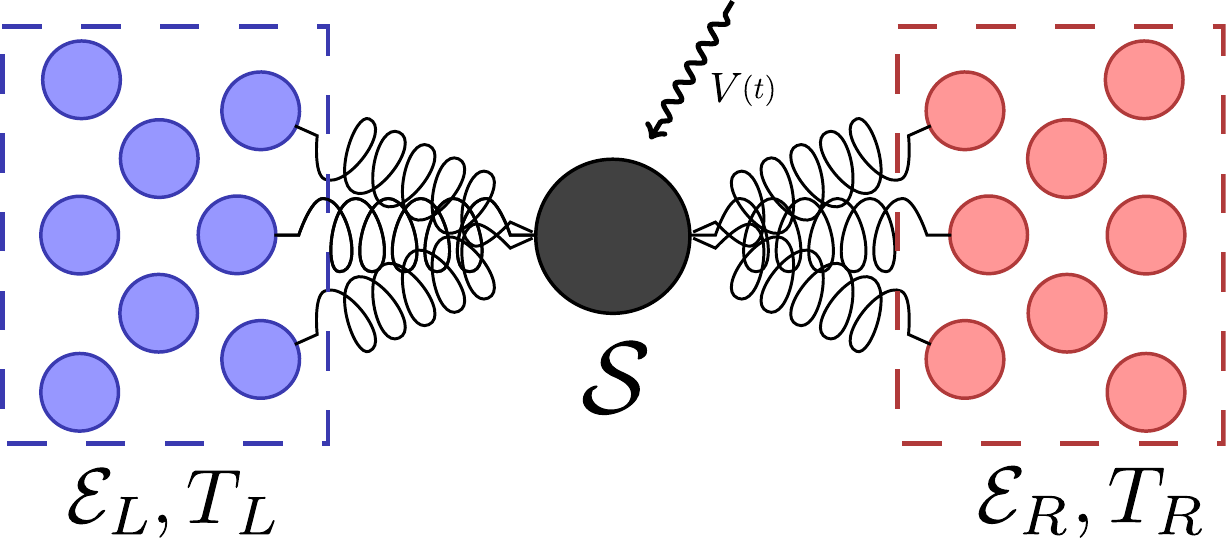}
    \caption{A parametric oscillator $\sys$ driven by $V \left( t \right)$ is coupled with two environments $\env_{L}$ and $\env_{R}$ at temperatures $T_{L}$ and $T_{R}$, respectively.}
    \label{sistema}
\end{center}
\end{figure}

We consider a generalization of the usual quantum Brownian motion model (QBM) \cite{huPazZhang,caldeiraLeggett} that was presented and solved in Ref. \cite{aguilarFreitasPaz}. It consists of a parametric oscillator, which we will refer to as system $\sys$, coupled with an environment $\env$ formed by $N$ independent oscillators. The environment is then divided into two pieces $\env_{L}$ and $\env_{R}$ by preparing them in different thermal states with temperatures $T_{L}$ and $T_{R}$, respectively. This model represents the physical situation shown in Fig. \eqref{sistema}.
Since the solution of this model has been treated in detail in \cite{aguilarFreitasPaz}, here we will only present a brief outlook of it. The dynamics are governed by the total Hamiltonian $H_{T} = H_{\sys} \otimes \mathbb{1}_{\env} + \mathbb{1}_{\sys} \otimes H_{\env} + H_{\sys,\env}$. The system's Hamiltonian is
$H_{\sys}= p^{2}/2m+ m V(t) x^{2}/2$ while the environmental and interaction terms are, respectively, $H_{\env} = \sum_{i} \left( p_{i}^{2}/2m_{i} + m_{i} \omega_{i}^{2} q_{i}^{2}/2 \right)$ and $H_{\sys,\env} = x \sum_{i} \lambda_{i} q_{i}$. The solution of the Heisenberg equations of motion is
\begin{equation}
    \begin{aligned}
    & q_{i} = q_{i}^{h} + K_{ij}^{ \left( 1 \right)} \ast q_j^{h} + K^{\left( 2 \right)}_{i} \ast x_{h},\\
    & x = x_{h} + K^{\left( 3 \right)}_{j} \ast q_{j}^{h},
    \end{aligned}
    \label{evolucionCoordenadas}
\end{equation}
where the notation $ F \ast f = \int_{0}^{t} dt^{\prime} F \left( t, t^{\prime} \right) f \left( t^{\prime} \right)$ is used. $q_{i}^{h} \left( t \right) = q_{i,0} \cos \left( \omega_{i} t \right) + p_{i,0} \sin \left( \omega_{i} t \right)/m_{i} \omega_{i}$ are the free Heisenberg operators of the environmental modes, with $q_{i,0}$ and $p_{i,0}$ Schr\"oedinger operators, and $x_{h} \left( t \right)$ is a dressed operator for $\sys$ that satisfies the linear equation
\begin{equation}
    \ddot{x}_{h}  + V_{R} \left( t \right) x_{h} + \gamma \ast \dot x_h = 0.
    \label{sistemaHomogenea}
\end{equation}
Above, $\gamma$ is the dissipation kernel $\gamma (t) = \int d  \omega I (\omega) \text{cos}(\omega t)/ m \omega$, and $V_{R} \left( t \right) = V \left( t \right) - \gamma \left( 0 \right)$ is the renormalized potential. The explicit expressions for the kernels $K^{\left( 1,2,3 \right)}$ can be found in \cite{aguilarFreitasPaz}, but it is sufficient to know that they are functionals of the Green function $G$ of Eq. \eqref{sistemaHomogenea}. Thus, knowing $G$ means solving the full Heisenberg equations. For a periodic driving with frequency $\omega_{d}$, $V \left( t \right) = \sum_{k} V_{k} e^{i k \omega_d t}$, the Green function $G$ can always be written as $G \left( t , t^{\prime} \right) = \textstyle{\sum_{k}} A_{k} \left( t - t^{\prime} \right)  e^{i k \omega_{d} t}$, where $A_{k}$ vanishes for negative arguments. For $G$ to be a Green's function of Eq. \eqref{sistemaHomogenea}, the Laplace transform of the functions $A_{k} (t)$, $\tilde{A}_{k} \left( s \right)$,  must satisfy a linear set of algebraic equations:
\begin{equation}
    \tilde{g}^{-1} \left( s + i k \omega_{d} \right) \tilde{A}_{k} \left( s \right) + \textstyle{\sum_{n \neq 0}} V_{n} \, \tilde{A}_{k - n} \left( s \right) = \delta_{k 0},
    \label{eqAk}
\end{equation}
where $\tilde{g}$ is the Laplace transform of the static Green function. The above system of equations can be simply solved by using a perturbative series expansion which is valid when the Fourier coefficients of the potential $\lvert V_{k} \rvert$ are small and the frequency of the driving $\omega_{d}$ is detuned from the parametric resonance (that is, from the renormalized frequency $\omega´_{r} = V_{0} - \gamma (0)$ \cite{landau}). In that case, the solution of \eqref{eqAk} satisfies the following recurrence relation:
\begin{equation}
    A_{k}^{\left( m \right)} \! \left( s \right) = \tilde{g} \left( s + i k \omega_{d} \right) [ \delta_{k 0} - \textstyle{\sum_{n \neq 0}} V_{n} \, \tilde{A}_{k - n}^{\left( m - 1 \right)} \left( s \right) ],
    \label{potenciasV}
\end{equation}
for $m\geq 1$, with $A_{k}^{\left( 0 \right)} \left( s \right) = \tilde{g} \left( s + i k \omega_{d} \right) \, \delta_{k 0}$.

We will focus our attention on Gaussian initial states with first moments equal to zero. Since the total Hamiltonian is quadratic, states remain Gaussian for all times. Also, our results are independent of the specific form of the spectral density $I_{R,L} ( \omega)$ of the environments, but we will require that it is a smooth function of $\omega$ with a frequency-independent damping constant $\gamma_{0}$. That is, $I ( \omega) = \gamma_{0} \, f ( \omega )$ with $f$ smooth.

\section{The covariance matrix}
In order to make this paper self-contained, in this section we include some material from our previous work \cite{aguilarFreitasPaz}. We added new and better explanations to aid the understanding of the broader topics treated here.

Our goal is to study environmental correlations between two parts of the environment, so we use Eq. \eqref{evolucionCoordenadas} to compute all correlation functions between two environmental bands: one of them consists of oscillators whose frequencies are distributed around $\omega_i \in \env_{R}$ with a bandwidth $\Delta \omega$, and the other one is centered around $\omega_{j} \in \env_{L}$. All expressions will depend on the product $I(\omega)\Delta\omega$ which, for sufficiently small values of $\Delta\omega$, plays the role of an effective coupling strength between the reservoir band and $\sys$ (since $I(\omega)\Delta\omega\approx\lambda^2/m\omega$). With the correlation functions we build the two-mode covariance matrix as $\boldsymbol{\sigma}_{ab} (t) = \langle \{ z_{a} (t) , z_{b} (t) \} \rangle / 2$ where $\vec{z} = (q_{i}, p_{i}, q_{j}, p_{j})$, which determines all properties of Gaussian states. For the sake of simplicity we avoid writing all correlators here (in Appendix A we include the general form of the position and momentum correlators), but we can prove that, for a continuous environment (the continuous hypothesis is discussed in Ref. \cite{aguilarFreitasPaz}),
\begin{equation}
    \boldsymbol{\sigma} (t) = \boldsymbol{\nu}_{i} \otimes \boldsymbol{\nu}_{j} + \boldsymbol{\sigma}_{0} (t) + \boldsymbol{\sigma}_{osc} (t) + \boldsymbol{\sigma}_{lin} \times t,
    \label{covariance}
\end{equation}
where $\boldsymbol{\nu}_{i} \otimes \boldsymbol{\nu}_{j}$ is the initial thermal state of both bands, $\boldsymbol{\sigma}_{0} (t)$ depends on the initial state of $\sys$ and $\boldsymbol{\sigma}_{0} (t) \to 0$ exponentially fast as $t \to \infty$, $\boldsymbol{\sigma}_{osc} (t)$ oscillates in time with frequencies $\omega_{d}$ and its higher harmonics, and $\boldsymbol{\sigma}_{lin} \times t$ is linear in time. We note that the fact that $\boldsymbol{\sigma}_{0} (t) \to 0$ is not trivial and rest on the assumption that a stable stationary regime exists. The existence of such regime requires the energy pumped into $\sys$ to be dissipated by $\env$. This can be achieved for small driving amplitudes provided that $\omega_d$ is detuned from the parametric resonance.

We are interested in the physical processes that continuously create correlations beyond the transient regime. Therefore, we will only keep the terms that do not equal zero after averaging $\boldsymbol{\sigma}$ over a driving cycle in the long-time limit. Thus, we will work with
\begin{equation}
    \boldsymbol{\sigma}_{av} (t) = \boldsymbol{\nu}_{i} \otimes \boldsymbol{\nu}_{j} + \boldsymbol{\sigma}_{lin} \times t.
    \label{covarianceAV}
\end{equation}
The linear term in time, $\boldsymbol{\sigma}_{lin}$, is a block diagonal matrix (that is, the cross-correlators are zero) unless the center frequencies of the bands satisfy a precise relation: either $\omega_{i} + \omega_{j} = k \omega_{d}$ or $\omega_{j} = \omega_{i} + k \omega_{d}$, with $k \in \mathbb{Z}$. This is the signature of the two main processes that creates correlations between the bands. In order to understand their nature, we compute the energy $E_{i}$ stored in the band $i \in \env_{R}$, which varies due to the same two processes. $E_{i}$ is proportional to the sum of the two first diagonal correlators of $\boldsymbol{\sigma}_{av}$ ($E_{i} \propto  \boldsymbol{\sigma}_{av,11} + \boldsymbol{\sigma}_{av,22}$, see Appendix A for a brief derivation) and has a simple expression: $E_i (t) = \left[ 1/2 + n_{R} \left( \omega_{i} \right) \right] \omega_{i} + \dot{\mathcal{Q}}_{i} \times t$, with
\begin{equation}
	\begin{aligned}
		&\frac{\dot{\mathcal{Q}}_{i}}{\Delta\omega}=
 		 \sum_{k} \sum_{\alpha = R,L} \omega_i \left\{ \Theta \left( \omega_{i, k} \right) p_{R,\alpha}^{(k)} \left( \omega_{i} \right) \left[ n_{\alpha} \left( \omega_{i,k} \right) - n_{R} \left( \omega_{i} \right) \right] \right. \\
        & \left. + \Theta \left(-\omega_{i, k} \right) p_{R,\alpha}^{(k)} \left( \omega_{i} \right) \left[ n_{\alpha} \left( \left\lvert \omega_{i,k} \right\rvert \right) + n_{R} \left( \omega_i \right) + 1 \right] \right\},
    \end{aligned}
    \label{heatCurrent}
\end{equation}
where $\omega_{i,k} = \omega_{i} - k \omega_{d}$, $\Theta$ is the step function, $n_{\alpha} \left( \omega \right)$ is the Planck distribution with temperature $T_{\alpha}$, and $p_{R,\alpha}^{(k)} \left( \omega_{i} \right) = \pi \,  I_{R} \left( \omega_i \right) I_{\alpha} \left( \left\lvert \omega_{i, k} \right\rvert \right) \lvert \tilde{A}_{k} \left( i \omega_{i,k} \right) \rvert^{2}/2m^2$ is the probability that mode $\omega_{i}$ in $\env_{R}$ interacts with mode $\omega_{i, k}$ in $\env_{\alpha}$ through $\sys$ ($I_{\alpha}$ is the spectral density of $\env_{\alpha}$). In Eq. \eqref{heatCurrent} we can see the two main processes that generate correlations between the bands. In the long-time regime the energy stored in each band varies due to these two processes and their relative importance depends on the temperatures $T_{R}$ and $T_{L}$. As the first term on the right-hand side of Eq. \eqref{heatCurrent} shows, at nonzero environmental temperature, the resonant absorption (or emission) from (or into) the driving field can transport an excitation in a mode with frequency $\omega_{i}$ to a mode with frequency $\omega_{i} - k \omega_{d}$. This process is associated with the classical heat current that flows from one environment to another and, naturally, it is not present at zero temperature where there are no excitations to be transported around. Thus, at very low temperature, $E_{i}$ varies because of a different process. In the second term of Eq. \eqref{heatCurrent}, we can see that the energy of the driving is dumped into two modes whose frequencies add up to a multiple of $\omega_d$. This splitting of the driving energy between two modes generates entanglement between them, and is interpreted as the nonresonant creation of a pair of excitations (one in each mode) \cite{aguilarFreitasPaz,freitasPaz17,freitasPaz18}. This second term is positive and therefore it is always associated with heating\textemdash more specifically, with the third law of thermodynamics.

We have seen that the only time-extensive correlations are generated between environmental bands that satisfy either $\omega_{i} + \omega_{j} = k \omega_{d}$ or $\omega_{j} = \omega_{i} + k \omega_{d}$. The former are correlated by the nonresonant creation of a pair of excitations, while the latter are correlated by the resonant transport of excitations. In following sections we will study the nature of the correlations generated by these two processes and their relation with entanglement. Before that, we note that in the time-extensive regime, $\boldsymbol{\sigma}_{av}$ is symplectically equivalent to the standard form (see Lemma I in Ref. \cite{duanGiedkeCirac})
\begin{equation}
    \boldsymbol{\sigma}_{sf} =
    \begin{pmatrix}
    a & 0 & c_{1} & 0 \\
    0 & a & 0 & c_{2} \\
    c_{1} & 0 & b & 0 \\
    0 & c_{2} & 0 & b
    \end{pmatrix}
    \label{standardForm}
\end{equation}
where $c_{1} = c_{2}$ for the bands correlated by the resonant transport of excitations and $c_{1} = -c_{2}$ for the ones correlated by the nonresonant creation of a pair of excitations. This is the subclass of squeezed-thermal states. Therefore the driving might be creating squeezed thermal states between the bands in the environment.

\section{The toolbox}
In order to study the cause and nature of intraenvironmental correlations in thermal machines, we need:
\begin{enumerate}[(i)]
    \setlength{\itemsep}{-1pt}
    \item a measure of the total amount of correlations,
    \item a measure of quantum correlations, and
    \item a measure of entanglement.
\end{enumerate}
For the first item, the immediate choice is the mutual information $\mathcal{I}$. For a bipartite system divided in $A$ and $B$, it is defined as $\mathcal{I} ( A : B) = S (A) + S(B) - S(AB)$, where $S$ is the von Neumann entropy. It is worth noticing that, whereas the mutual information for classical distributions is bounded from above by the entropy of either one of them, the mutual information for quantum systems can be as large as $2 \, \text{min} \{ S (A) , S(B) \}$ for entangled states (see Araki-Lieb inequality\textemdash Corollary of Theorem 2 in Ref. \cite{arakiLieb}), which reflects the existence of quantum correlations beyond the classical ones. In order to quantify these quantum correlations, we will use a measure called quantum discord (QD). QD was proposed \cite{ollivierZurek,hendersonVedral} in an effor to adress a series of developments in Refs. \cite{knillLaflamme,bennettDiVincenzoFuchs,horodeckiSendeSen} that challenged the belief that entanglement is the only form of quantum correlation. QD is defined by quantizing the mismatch between two classically equivalent measures of mutual information: $\overleftarrow{D} (A:B) = \mathcal{I} (A : B) - \mathcal{J} (A : B)$, where $\mathcal{I} (A : B) = H (A) + H(B) - H(A,B)$ and $\mathcal{J} (A : B) = H(A) - H(A \rvert B)$, with $H$ the Shannon entropy. The discrepancy between $\mathcal{I}$ and $\mathcal{J}$ in the quantum case occurs because, in quantum theory, the conditional entropy $H(A \rvert B)$ involves an specific choice of basis to perform a measurement on $B$ to infer the state of $A$. Thus, the quantization of $\mathcal{J}$ is not as straightforward as the one of $\mathcal{I}$, which only requires to replace the Shannon entropy for the von Neumann one. In order to do this, a minimization is performed over all positive operator valued measures (POVMs) corresponding to a measurement in $B$ so as to find the one that disturbs least the overall quantum state and that, at the same time, allows to extract the most information about $B$. Hence, the quantum analog of $\mathcal{J}$ is defined as
\begin{equation}
    \overleftarrow{\mathcal{J}} (A:B) = S(A) - \inf_{\{ \Pi_{n} \}} \left[ \sum_{n} \, p_{n} \, S( A | n)  \right],
\end{equation}
where $\{ \Pi_{n} \}$ is a POVM, $p_{n} = \text{tr} ( \rho_{B} \Pi_{n})$ is the probability of obtaining the result $n$, and $S( A | n)$ is the von Neumann entropy of the reduced state of $A$ after obtaining this result. Quantum discord has been shown to be a property held by almost all quantum states \cite{ferraroAolitaCavalcanti} (the set of states with zero quantum discord has measure zero and is nowhere dense) and has attracted considerable attention \cite{dattaShajiCaves,rodriguezrosarioModiKuah,pianiHorodeckiHorodecki,lanyonBarbieriAlmeida,freitasPaz12}. Initially defined only in finite dimensional systems, the concept of QD was extended to continuous-variable systems, specifically to the case of two-mode Gaussian states \cite{giordaParis,adessoDatta}. When the POVMs are restricted to the set of Gaussian measurements, it is called Gaussian quantum discord (GQD). Until recently it was thought to be an upper bound of the QD for continuous-variable systems but it has been proved that, in fact, QD and GQD are equal for Gaussian states \cite{pirandolaSpedalieriBraunstein}. Finally, as a measure of entanglement we will use the logarithmic negativity $E_{\mathcal{N}} = \text{log} \| \boldsymbol{\tilde{\rho}} \|_{1}$, where $\tilde{\rho}$ is the partially transposed density matrix. $E_{\mathcal{N}}$ quantifies the violation of the Peres-Horodecki criterion and it is monotone under local operations and classical communication \cite{adessoIlluminati}. In the case of two-mode Gaussian states, this criterion is a necessary and sufficient condition for the composite system to be separable.

Gaussian states are completely determined by their covariance matrix $\boldsymbol{\sigma}$, so it is not surprising that all three quantities discussed above can be computed using only $\boldsymbol{\sigma}$. For example, the logarithmic negativity is computed as $E_{\mathcal{N}} = \text{max} \{ 0 , - \text{ln} (2 \tilde{\lambda}_{-}) \}$, where $\tilde{\lambda}_{-}$ is the lowest symplectic eigenvalue of the covariance matrix $\boldsymbol{\tilde{\sigma}}$ corresponding to the partially transposed density matrix $\boldsymbol{\tilde{\rho}}$ (which differs from $\boldsymbol{\sigma}$ by just a sign flip in the off-diagonal block matrices). Expressions for the mutual information and quantum discord in terms of $\boldsymbol{\sigma}$ can be respectively found in Ref. \cite{serafiniIlluminatiSiena} and in Ref. \cite{adessoDatta}.

\section{Correlations between environmental bands}
In this section we present a study of classical and quantum correlations between two environmental bands which are respectively centered around frequencies $\omega_{i} \in \env_{R}$ and $\omega_{j} \in \env_{L}$, in increasing order of quantumness. We will use the toolbox described in the previous section. First, we will compute the mutual information $\mathcal{I}$ for the total correlations, second, the quantum discord $\overleftarrow{D}$, which measures the quantumness of correlations, and third, as a measure of entanglement, the logarithmic negativity $E_{\mathcal{N}}$. We will analyze the relationship between these quantities and present simple analytic expressions which are valid in the limit of small driving and weak coupling ($\gamma_{0} / \omega_{i,j} \ll 1$). We will examine and compare two distinct physical processes which, as discussed above, are the main sources of generation of time-extensive correlations in the long-time regime: the non-reasonant creation of a pair of excitations, satisfying the condition $\omega_{i} + \omega_{j} = k \omega_{d}$, and the resonant transport of excitations, with $\omega_{j} = \omega_{i} + k \omega_{d}$.

Our formulas are written in terms of only three physically meaningful symplectic invariants. These are: the individual purities $\mu_{i} = \text{tr} (\rho_{\alpha}^{2}) = 1 / 2 \sqrt{\text{det} ( \boldsymbol{\alpha})}$ and $\mu_{j} = 1 / 2 \sqrt{\text{det} ( \boldsymbol{\beta} )}$, and the determinant $\Gamma (t) = 4 \, \text{det} ( \boldsymbol{\gamma} )$, which is a measure of the generation of cross correlations (see below). Here, $\boldsymbol{\alpha}$, $\boldsymbol{\beta}$ and $\boldsymbol{\gamma}$ are the three $2 \times 2$ submatrices in the two-mode covariance block matrix $\boldsymbol{\sigma}$. If we write $\boldsymbol{\sigma}$ in its standard form as in Eq. \eqref{standardForm}, then $\boldsymbol{\alpha} = a \, \mathbb{1}$, $\boldsymbol{\beta} = b \, \mathbb{1}$, and $\boldsymbol{\gamma} = \text{diag} (c_{1} , c_{2})$.

Although the individual purities depend on time (they decrease as time goes by), in the weak-coupling limit their value is mostly determined by the temperature of the environments. That is, $\mu_{i,j} \to 1^{-}$ indicates low temperature whereas $\mu_{i,j} \to 0^{+}$ indicates high temperature. On the other hand, $\Gamma (t)$ is a quadratic function of time, $\Gamma_{\pm} (t) = \Gamma_{\pm} \times t^{2}$ (from now on, the plus sign will indicate the nonresonant case and the minus sign, the resonant one). As we mentioned above, $\Gamma_{\pm} (t)$ can be interpreted as the generator of cross-correlations: pair creation or transport of excitations, depending on the case. This is because the Kullback-Leibler divergence between the Wigner function of $\rho_{\boldsymbol{\sigma}}$ and the Wigner function of the product of the marginals $\rho_{\boldsymbol{\alpha}} \otimes \rho_{\boldsymbol{\beta}}$ can be written as
\begin{equation}
    D_{KL} ( W_{\rho_{\boldsymbol{\sigma}}} || W_{\rho_{\boldsymbol{\alpha}} \otimes \rho_{\boldsymbol{\beta}}} ) \simeq \mu_{i} \, \mu_{j} \, \lvert \Gamma_{\pm} ( t) \rvert
    \label{KL}
\end{equation}
(see Ref. \cite{adessoGirolamiSerafini}). $D_{KL}$ measures the phase-space distinguishability between $\rho_{\boldsymbol{\sigma}}$ and $\rho_{\boldsymbol{\alpha}} \otimes \rho_{\boldsymbol{\beta}}$. As we can see from Eq. \eqref{KL}, while the purities decrease with time, $\lvert \Gamma_{\pm} ( t) \rvert$ increases quadratically making $\rho_{\boldsymbol{\sigma}}$ more and more distinguishable from the completely uncorrelated state $\rho_{\boldsymbol{\alpha}} \otimes \rho_{\boldsymbol{\beta}}$. Thus, $\Gamma_{\pm} (t)$ is indeed responsible for generating correlations.

In Appendix B we include explicit expressions for the individual purities in Eq. \eqref{muExplicit} and for $\Gamma_{\pm}$ in Eqs. \eqref{gmenosExplicit} and \eqref{gmExplicit}, respectively. In Appendix C we include a brief derivation of the equations presented below.

\subsection{Mutual information}
We begin our study of correlations by computing the mutual information between the bands. For both the resonant and nonresonant cases, $\mathcal{I}$ has the same simple form
\begin{equation}
    \mathcal{I}_{\pm} ( \boldsymbol{\sigma}_{av} ) \simeq f_{\pm} ( \mu_{i} , \mu_{j} ) \, \lvert \Gamma_{\pm} (t) \rvert
    \label{mutualInformation}
\end{equation}
with
\begin{equation}
f_{\pm} ( \mu_{i} , \mu_{j} ) = \mu_{i} \mu_{j} [ \text{atanh} (\mu_{i}) \pm \text{atanh} (\mu_{j}) ] / ( \mu_{i} \pm \mu_{j} ).
\end{equation}
Although Eq. \eqref{mutualInformation} may give the impression that mutual information grows quadratically in time like $\, \lvert \Gamma_{\pm} (t) \rvert$, this is not the case. The reason is that, while $\, \lvert \Gamma_{\pm} (t) \rvert$ increases, $f_{\pm}$ is a monotonically decreasing function of the individual purities. In fact, for the resonant case it can be shown (see Appendix B) that
\begin{equation}
    \lvert \Gamma_{-} (t) \rvert \leq (1 - \mu_{i}) (1 - \mu_{j}) / \mu_{i} \mu_{j},
    \label{ineq}
\end{equation}
and, consequently, the mutual information is bounded: $\mathcal{I}_{-} \leq 1$. Thus, the quadratic growth in time must eventually stop and $\mathcal{I}_{-}$ reaches a saturation value. It seems that the generation of correlations by means of the transport of excitations cannot increase the mutual information between the bands to arbitrarily large values. On the contrary, in the nonresonant case the bound in Eq. \eqref{ineq} does not hold. This is a natural result taking into account that this process continually creates correlations in time. However, it exists a regime in which the mutual information $\mathcal{I}_{\pm}$ grows relatively quadratically in time. For not so long-times such that the individual purities $\mu_{i,j}$ stay approximately constant, $f_{\pm}$ stays constant too and the mutual information grows as $\, \lvert \Gamma_{\pm} (t) \rvert$.

\begin{figure}[htp]
    \begin{center}
    \includegraphics[scale=.50]{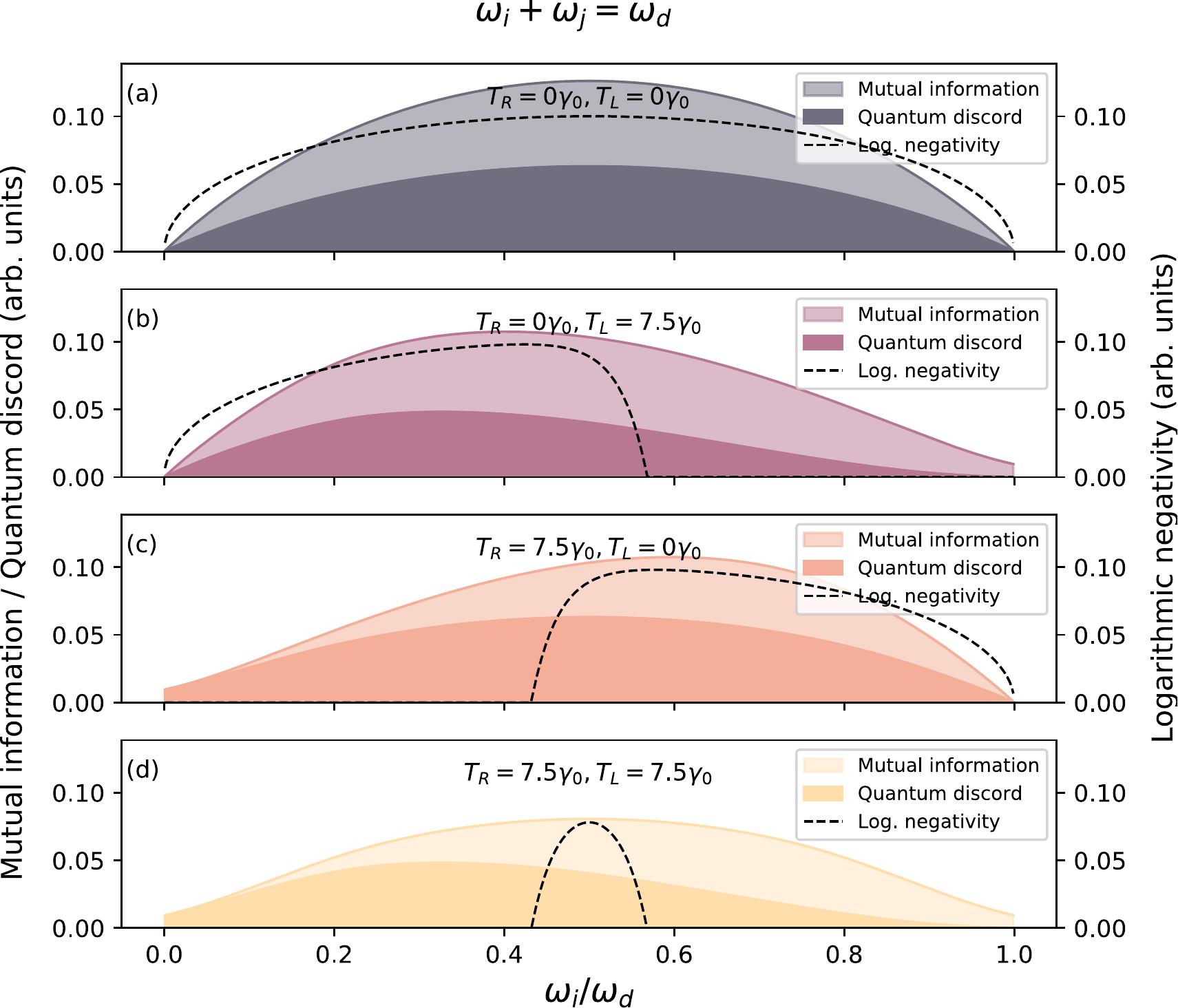}
    \caption{Dependence of the mutual information, quantum discord and logarithmic negativity on the frequency $\omega_{i}$ for the nonresonant case, with $\omega_{i} + \omega_{j} = \omega_{d}$. (a) When both environments are at zero temperature, entanglement is maximized and $\protect\overleftarrow{D}_{+} / \mathcal{I}_{+} \to 1/2$. (b,c) Temperature is raised in one of the environments to $T = 7.5 \gamma_{0}$, while the other one is kept at $T=0$. This illustrates the asymmetry of the quantum discord. (d) Both temperatures are raised to $T = 7.5 \gamma_{0}$ and the only entanglement left is present at the center frequencies. Here, $V (t) = \omega_{r}^{2} + V \text{cos} (\omega_{d} t)$ and $I ( \omega ) = 2 m \gamma_{0} \omega \Lambda^{2} / \pi (\omega^{2} + \Lambda^{2})$. The plots are normalized using $E_{0}= \gamma_{0} \Delta \omega V t / \omega_{r}^{3}$ (we plot $\mathcal{I}_{+}/E_{0}^{2}$, $\protect\overleftarrow{D}_{+} / E_{0}^{2}$ and $E_{\mathcal{N}}/E_{0}$). The parameters used are $\omega_{d} = \omega_{r} / \sqrt{11}$, $\omega_{r} = 800\gamma_0$, $V=\omega_r^2/32$, $m=10m_i$, $t=20\gamma_0$ and $\gamma_0 = 0.005$.}
    \label{nonResonant}
\end{center}
\end{figure}

\begin{figure}[htp]
    \begin{center}
    \includegraphics[scale=.60]{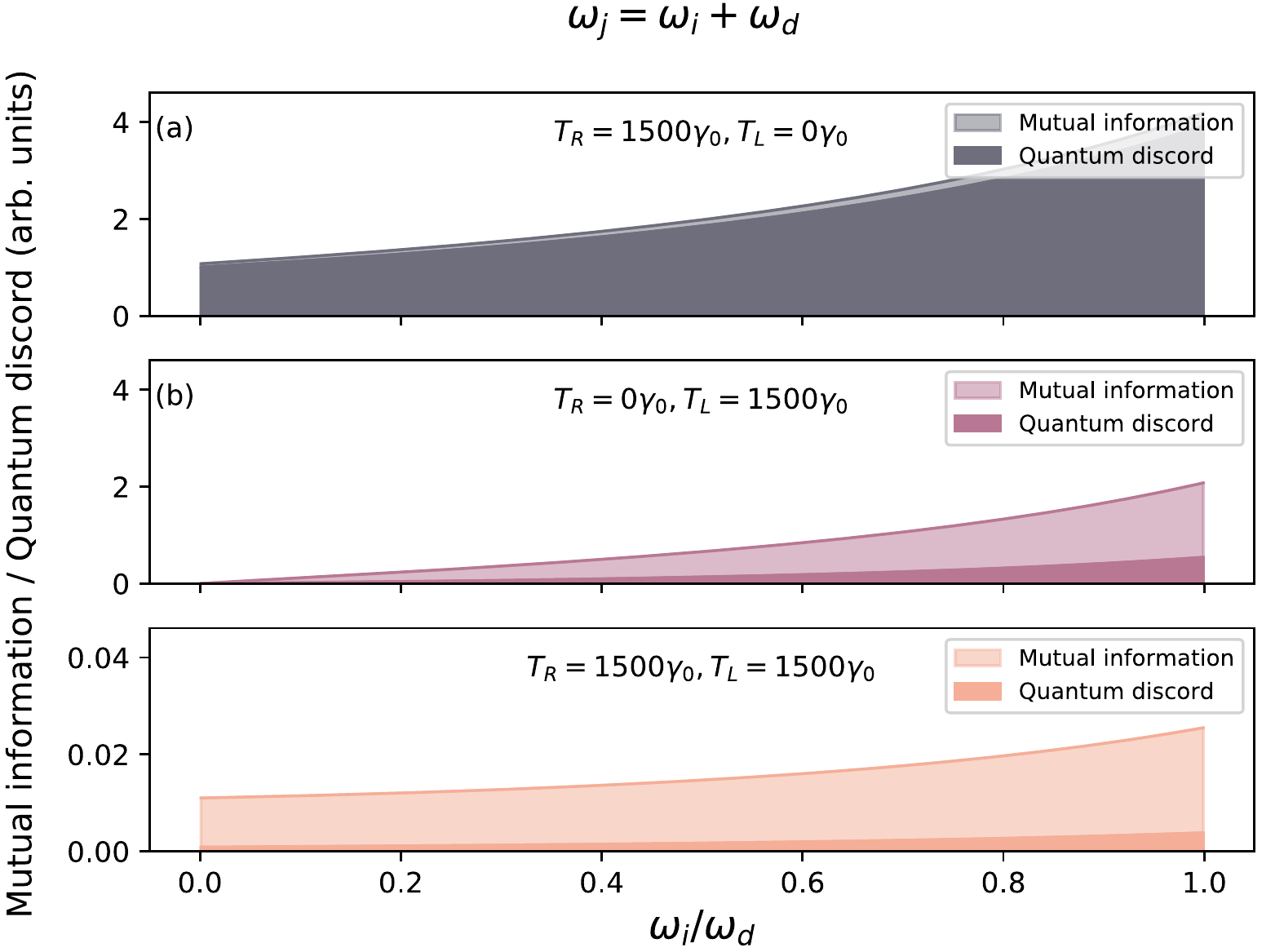}
    \caption{Dependence of the mutual information and quantum discord on the frequency $\omega_{i}$ for the resonant case, with $\omega_{j} = \omega_{i} + \omega_{d}$. (a,b) Temperature is raised in one of the environments to $T = 1500 \gamma_{0}$, while the other one is kept at $T=0$. This illustrates the limits shown in Eqs. \eqref{discordR1} and \eqref{discordR2}, respectively. (c) Both temperatures are raised to $T = 7500 \gamma_{0}$, showing that $\mathcal{I}_{-}$ becomes vanishingly small when occupation numbers are similar (note that the scale of the vertical axis is 100 times smaller than the previous plots). Here, $V (t) = \omega_{r}^{2} + V \text{cos} (\omega_{d} t)$ and $I ( \omega ) = 2 m \gamma_{0} \omega \Lambda^{2} / \pi (\omega^{2} + \Lambda^{2})$. The plots are normalized using $E_{0}= \gamma_{0} \Delta \omega V t / \omega_{r}^{3}$ (we plot $\mathcal{I}_{-}/E_{0}^{2}$ and $\protect\overleftarrow{D}_{-} / E_{0}^{2}$). The parameters used are $\omega_{d} = \omega_{r} / \sqrt{11}$, $\omega_{r} = 800\gamma_0$, $V=\omega_r^2/32$, $m=10m_i$, $t=20\gamma_0$ and $\gamma_0 = 0.005$.}
    \label{resonant}
\end{center}
\end{figure}

With this in mind, Eq. \eqref{mutualInformation} has a clear physical interpretation: mutual information increases due to the creation of cross-correlations between the bands by $\Gamma_{\pm} (t)$, but this increment is countered by the influence of the thermal agitations in the environment, which are represented by $f_{\pm}$. Certainly, as time passes or temperature increases the bands correlate progressively more with the rest of the environment, decreasing its mutual dependence. Furthermore, since $f_{+} < f_{-} $, the effects of the thermal agitations in the environment are more detrimental to the bands that are related by the nonresonant pair creation than to the ones that are related by the more ``classical'' resonant transport of excitations. Mutual information is plotted in Figs. \ref{nonResonant} and \ref{resonant} for different environmental temperatures, in the nonresonant and resonant cases, respectively.

An interesting feature can be seen from the form of $\Gamma_{-}$:
\begin{equation}
    \Gamma_{-} \propto \lvert n_{R} ( \omega_{i} ) \tilde{A}_{k} ( i \omega_{i} ) - n_{L} ( \omega_{j} ) \tilde{A}_{-k}^{\ast} ( i \omega_{j} ) \rvert^{2}
    \label{gmProp}
\end{equation}
(the complete expression can be found in Eq. \eqref{gmExplicit} in Appendix B). In the resonant case, the transport of excitations can go both ways depending if a quantum of energy is either absorbed or emitted from the driving field. Because excitations act as identical quasiparticles, an interference effect takes place and the correlations generated by the absorption tend to cancel the ones generated by the emission. Indeed, the case where no excitations have been transported is indistinguishable from the one where the same number of excitations have been transferred in both directions. In particular, when the bands have the same occupation number (i.e., $n_{R} ( \omega_{i} ) = n_{L} ( \omega_{j} )$), absorption and emission are equally likely and mutual information tends to vanish. This is reflected in the fact that $\Gamma_{-}$ becomes vanishingly small, as can be seen in Eq. \eqref{gmProp}. This is illustrated in Fig. \ref{resonant}(c) (note the scale of the vertical axis).

Seeing that the pair creation mechanism is the dominant process at lower temperatures, physical intuition tells us that the mutual information between the bands correlated by it must be bigger than the ones correlated by the resonant transport of excitations. This is confirmed by an analysis of $ \Gamma_{\pm} $. As opposed to the nonresonant case, in the resonant case, $\Gamma_{-} \to 0$ as $T \to 0$ (see  Eq. \eqref{gmProp}). At zero environmental temperature the bands are not populated and, therefore, the driving cannot transport excitations from one into the other.

\subsection{Quantum discord}
In this section we will compute the ratio $\overleftarrow{D}/\mathcal{I}$ which represents the amount of quantum correlations among the totality of correlations. The result is
\begin{equation}
    \overleftarrow{D}_{\pm} ( \boldsymbol{\sigma}_{av} ) / \mathcal{I}_{\pm} ( \boldsymbol{\sigma}_{av} ) \simeq 1 - g_{\pm} (\mu_{i} , \mu_{j}),
    \label{discord}
\end{equation}
with
\begin{equation}
    g_{\pm} (\mu_{i} , \mu_{j}) = \frac{1}{1 + \mu_{j}} \left( 1 \pm \frac{\mu_{j}}{\mu_{i}} \right) \left[ 1 \pm \frac{\text{atanh} (\mu_{j})}{\text{atanh} (\mu_{i})} \right]^{-1}.
\end{equation}
We note that $g_{\pm}$ is such that $0 \leq g_{\pm} \leq 1$ and it is a monotonically increasing function of the purities. Since $g_{+} > g_{-} $, it follows that $\overleftarrow{D}_{+}/\mathcal{I}_{+} < \overleftarrow{D}_{-}/\mathcal{I}_{-}$. That is, the fraction of quantum correlations in the mutual information between the bands correlated by the nonresonant pair creation is less than the fraction between the bands correlated by the resonant transport of excitations. Nevertheless, it is possible for the quantum discord in the nonresonant case to be greater than the one in the resonant one (i.e., $\overleftarrow{D}_{+} > \overleftarrow{D}_{-}$). For example, that is the case in the low temperature limit where $\mathcal{I}_{-} \to 0$ but $\mathcal{I}_{+} \nrightarrow 0$. As a consequence of the mutual information in the resonant case being bounded, the quantum discord is bounded too: $\overleftarrow{D}_{-} \leq 1$. This result was previously obtained in Refs. \cite{adessoDatta, giordaParis} when studying quantum discord in squeezed thermal states. Thus, it is consistent with the idea that that is the type of state generated by the resonant process. Now we will provide some examples that illustrate the previous results in two different regimes.

We will analyze the case where environmental temperatures are similar (a more precise statement would be $\mu_{i} \simeq \mu_{j}$, but in the limits of high and low temperatures both conditions are equivalent). We will express all quantities in terms of $\bar{\mu} = (\mu_{i} + \mu_{j})/2$ and $\Delta \mu = (\mu_{i} - \mu_{j})/2$, and we will work up to first order in $\lvert \Delta \mu \rvert / \bar{\mu} \ll 1$. In this case, the ratio $\overleftarrow{D}/\mathcal{I}$ is
\begin{equation}
    \overleftarrow{D}_{+} ( \boldsymbol{\sigma}_{av} ) / \mathcal{I}_{+} ( \boldsymbol{\sigma}_{av} ) \simeq \frac{\bar{\mu}}{1 + \bar{\mu}}
    \label{discordLowNR}
\end{equation}
and
\begin{equation}
    \overleftarrow{D}_{-} ( \boldsymbol{\sigma}_{av} ) / \mathcal{I}_{-} ( \boldsymbol{\sigma}_{av} ) \simeq 1 - \frac{1}{\bar{\mu}} ( 1 - \bar{\mu}) \, \text{atanh} (\bar{\mu}).
    \label{discordLowR}
\end{equation}
As we can see from Eqs. \eqref{discordLowNR} and \eqref{discordLowR} above, at high temperatures, when $\bar{\mu} \to 0^{+}$, both expressions go to zero (i.e., $\overleftarrow{D}/\mathcal{I} \to 0$). In the other limit, when $\bar{\mu} \to 1^{-}$, we have $\overleftarrow{D}_{+} / \mathcal{I}_{+} \to 1/2$ and $\overleftarrow{D}_{-} / \mathcal{I}_{-} \to 1$. For the nonresonant case, it means that half of the total correlations are quantum in origin. This can be interpreted as follows. For the quantum system, mutual information is bounded by $\mathcal{I}_{+} \leq 2 \, S_{\bar{\mu}}$ (Araki-Lieb inequality with $S_{\bar{\mu}}$ the entropy of an ``average band'' represented by $\bar{\mu}$), while for a classical system the mutual information satisfies $\mathcal{I}_{+} \leq S_{\bar{\mu}}$. Therefore, the above limit ($\overleftarrow{D}_{+} / \mathcal{I}_{+} \to 1/2$) seems to indicate that the pair creation mechanism at zero temperature saturates the Araki-Lieb inequality. Thus, $\mathcal{I}_{+} \to 2\, S_{\bar{\mu}}$, and, qualitatively, $0 < \mathcal{I}_{+} \leq S_{\bar{\mu}}$ corresponds to classical correlations and $S_{\bar{\mu}} < \mathcal{I}_{+} \leq 2 \, S_{\bar{\mu}}$ to quantum ones. This is shown in Fig. \ref{nonResonant} (a). On the other hand, in the resonant case the situation is different. When $\bar{\mu} \to 1^{-}$, the mechanism creating correlations tends to vanish (remember that in this case, $\mathcal{I}_{-} \ll \mathcal{I}_{+}$) but the small amount of correlations that are present will have nonzero quantum discord and satisfy $\overleftarrow{D}_{-} \simeq \mathcal{I}_{-}$. As time passes or temperature increases, classical correlations start to develop and the quantum discord begins to decrease.

Since the definition of the quantum discord is based on measurements over one of the subsystems, it is naturally asymmetric with respect to an $\{ i , R \} \leftrightarrow \{ j , L \}$ interchange. It is interesting to explore how this asymmetry manifests itself in terms of the temperatures of the environments. For example, let us use the resonant case (the nonresonant one is analog), and suppose one of the environments is at low temperature and the other one is at an arbitrary (but higher) one. If $\mathcal{E}_{L}$ is the one at low temperature, then Eq. \eqref{discord} can be approximated as
\begin{equation}
    \overleftarrow{D}_{-} ( \boldsymbol{\sigma}_{av} ) / \mathcal{I}_{-} ( \boldsymbol{\sigma}_{av} ) \simeq 1 - \frac{\mu_{j}}{2 \, \text{atanh} (\mu_{j})}.
    \label{discordR1}
\end{equation}
Equation \eqref{discordR1} shows that most correlations are quantum ones in that limit. In contrast, if $\mathcal{E}_{R}$ is the one at low temperature, we have
\begin{equation}
    \overleftarrow{D}_{-} ( \boldsymbol{\sigma}_{av} ) / \mathcal{I}_{-} ( \boldsymbol{\sigma}_{av} ) \simeq 2 \, \frac{\mu_{j}}{\mu_{i}}.
    \label{discordR2}
\end{equation}
Now if we increase the temperature of $\mathcal{E}_{L}$, quantum discord reaches its minimum value: all correlations are classical. As we can see from the results above, the quantumness of the correlations measured by the quantum discord highly depends on which system is being observed. This difference is shown in Figs. \ref{nonResonant} (b) and (c) for the nonresonant case, and Figs. \ref{resonant} (a) and (b) for the resonant one (in this case we explicitly show the limits of Eqs. \eqref{discordR1} and \eqref{discordR2}).

\subsection{Entanglement}
The generation of entanglement in the nonresonant case was addressed in our previous work \cite{aguilarFreitasPaz}. Here we rederive its main equations but this time written in terms of symplectic invariants. We expand on its interpretation and the connection between entanglement, the phase-space entropy and the generation of correlations.

In the resonant case we have $\text{det} (\boldsymbol{\gamma}) > 0$, and therefore the logarithmic negativity is zero. That is, bands with center frequencies such that $\omega_{j} = \omega_{i} + k \omega_{d}$ are not entangled. Therefore, we focus our attention on the bands correlated by the pair creation mechanism (i.e, such that $\omega_{i} + \omega_{j} = k \omega_{d}$). We can show that these bands are entangled and the logarithmic negativity $E_{\mathcal{N}}$ is the maximum between zero and
\begin{equation}
    \text{E}_{\mathcal{N}} (t) \simeq - S_{ij} + \Gamma_{\mathcal{N}} \times t
    \label{entanglement}
\end{equation}
\noindent where $\Gamma_{\mathcal{N}} = ( \mu_{i} + \mu_{j} ) e^{- 2 S_{ij}} \sqrt{\lvert \Gamma_{+} \rvert} / 2 \mu_{i} \mu_{j}$ and $S_{ij} = \text{ln} \left[ \left( \mu_{i}^{2} + \mu_{j}^{2} \right) / 2 \mu_{i}^{2} \mu_{j}^{2} \right] / 2 \geq 0$. We note that, just like it happens with the mutual information and the quantum discord, $\text{E}_{\mathcal{N}}$ is not linear in time as Eq. \eqref{entanglement} may suggest. Entanglement production is time-extensive as long as the individual purities stay approximately constant. From Eq. \eqref{entanglement} we can see that the creation of entanglement is a competition between the generation of correlations by the pair creation mechanism and the thermal agitations of the environment that try to destroy them. These thermal agitations have two distinct effects on the logarithmic negativity. First, they reduce the rate of generation $\Gamma_{\mathcal{N}}$ ($\Gamma_{\mathcal{N}} \leq \sqrt{\lvert \Gamma_{+} \rvert}$). $\Gamma_{\mathcal{N}}$ is a function of the individual purities and it reaches its maximum value when $\mu_{i,j} \to 1^{-}$ ($\Gamma_{\mathcal{\mathcal{N}}} \to \sqrt{\lvert \Gamma_{+} \rvert}$). And second, they provide a threshold $S_{ij}$ that the pair creation mechanism has to overcome to entangle the bands. Logarithmic negativity is plotted alongside mutual information and quantum discord in Figs. \ref{nonResonant} (a)-(d) for different environmental temperatures. Note that, as temperature is increased, entanglement disappears from the side that corresponds to the lower frequencies in the higher temperature environment.

It is worth noticing that $S_{ij}$ is obtained from the Shannon entropy of the phase-space Wigner distributions corresponding to $\rho_{\boldsymbol{\alpha}}$ and $\rho_{\boldsymbol{\beta}}$. Thus, it is a measure of the disorder of the bands. In fact, it can be written as
\begin{equation}
    S_{ij} = \bar{S}_{2} + \text{ln} \left[ \text{cosh} \left( \Delta S_{2} \right) \right] / 2,
    \label{Sij}
\end{equation}
where $\bar{S}_{2} = \left[ S_{2} (\boldsymbol{\alpha}) + S_{2} (\boldsymbol{\beta}) \right] / 2$ and $\Delta S_{2} = S_{2} (\boldsymbol{\alpha}) - S_{2} (\boldsymbol{\beta})$, with $S_{2} (\rho) = - \text{ln} [ \text{tr} (\rho^{2}) ]$ the Renyi-2 entropy. For Gaussian states, $S_{2}$ coincides with the Shannon entropy of the corresponding phase-space Wigner distribution (up to an additive constant) \cite{adessoGirolamiSerafini}. As Eq. \eqref{Sij} shows, not only the entropy of the bands affects the generation of entanglement, but their difference too. For example, since for high temperatures $S_{2} \sim \text{ln} (T)$, it is harder to entangle two bands at different temperatures than two at the same one. $S_{ij}$ in turn fixes a latency time $t_{ent} = S_{ij} / \Gamma_{\mathcal{N}}$, which is the time it takes for the bands to be entangled.

As we previously mentioned, the generator of intra environmental correlations is $\Gamma_{\pm} (t)$ (see Eq. \eqref{KL}). In view of the above results, in the nonresonant case, we can rewrite this generator as
\begin{equation}
    \lvert \Gamma_{+} (t) \rvert = \frac{\lvert \Gamma_{+} \rvert}{\Gamma_{\mathcal{N}}^{2}} [ E_{\mathcal{N}} (t) + S_{ij}]^{2} \qquad t > t_{ent}.
    \label{gammaNREnt}
\end{equation}
Equation \eqref{gammaNREnt} shows that, from the moment the bands are entangled onwards, the pair creation mechanism can be interpreted as having the dual effect of increasing that entanglement and generating entropy in the environment. It must be emphasized that, since $S_{ij}$ is mostly constant through time, the predominant effect is the production of entanglement. As a consequence of Eq. \eqref{gammaNREnt}, the mutual information in Eq. \eqref{mutualInformation} and the quantum discord in Eq. \eqref{discord} can be cast in terms of the logarithmic negativity and, for fixed $E_{\mathcal{N}}$, both are nonmonotonic functions of the purities, in accordance with previous results \cite{giordaParis}. We also note that the second term in Eq. \eqref{entanglement}, which represents the generation of entanglement, is proportional to the number of entangled pairs $\Delta \omega$, as it should be since the logarithmic negativity is additive. This is in opposition to the mutual information in Eq. \eqref{mutualInformation} (and the quantum discord), which is proportional to $\Delta \omega^{2}$ and not additive.

As we mentioned in the beginning of this section, bands such that $\omega_{j} = \omega_{i} + k \omega_{d}$ are not entangled. That is, the resonant process cannot create entanglement. Nevertheless, it may play a role in the entanglement present between bands correlated by the nonresonant process. Let us suppose the excitation in band $i$ that the driving absorbs to be transported to band $j$ was previously put there by the pair creation mechanism. In that case, there must exist a $j^{\prime}$ band such that $\omega_{i} + \omega_{j^{\prime}} = k^{\prime} \omega_{d}$. Thus, the $j$ and $j^{\prime}$ bands are now related by $\omega_{j} + \omega_{j^{\prime}} = ( k + k^{\prime} ) \omega_{d}$, meaning that they are, in fact, entangled. From this we conclude that it is possible for the resonant process to swap the entanglement created by the nonresonant process to other bands. This should be a higher order effect in the damping constant $\gamma_{0}$ and thus not visible in our expressions above.

\section{Conclusions}
In this paper we presented a complete analysis of the origin and nature of the time-extensive correlations that are present in the reservoirs of a thermal machine (which we divided in the so-called bands). By studying a generalization of the usual QBM model that includes a time-dependent driving enforced on the system, we were able to show that there are only two processes that generate intraenvironmental correlations beyond the transient regime: the resonant transport of excitations, responsible for the classical heat flow, and the nonresonant pair creation, linked to the third law of thermodynamics. We would like to stress the most important results:
\begin{enumerate}[(i)]
    \setlength{\itemsep}{-1pt}
    \item There is a regime where mutual information and quantum discord between the bands grow approximately quadratically in time. This happens as long as the individual purities stay approximately constant. Furthermore, in the resonant case (with center frequencies such that $\omega_{j} = \omega_{i} + k \omega_{d}$) both quantities are bounded from above (showing that this regime must eventually end).
    \item Quantum discord is always present between the bands, independently of the process that correlates them. As shown in Eq. \eqref{discord}, the fraction of quantum correlations is negatively affected by the thermal agitations of the environment. This effect is worse on the bands correlated by the nonresonant process than the ones correlated by the resonant one.
    \item Entanglement is only present on the bands that are connected by the nonresonant process (with center frequencies such that $\omega_{i} + \omega_{j} = k \omega_{d}$). We showed that the logarithmic negativity grows approximately linear in time and that the pair creation mechanism has to overcome a threshold $S_{ij}$ to entangle the bands that is a measure of the disorder of them (see Eq. \eqref{Sij}).
    \item The resonant process cannot create entanglement --- at most it can swap it. Since it can only transport excitations between bands, if the band on which the excitation is absorbed was already entangled with another one, it can swap that entanglement to the band on which the excitation is dumped. This is a higher order effect which was not included in our formulas above.
\end{enumerate}
Our results are consistent with previous studies of classical and quantum correlations in Gaussian states \cite{giordaParis,adessoDatta}. Using the inequality presented in Eq. \eqref{ineq} we showed that the quantum discord between the bands correlated by the resonant process is bounded from above. Additionally, using Eq. \eqref{gammaNREnt}, we showed that the mutual information (and the quantum discord) between the ones correlated by the nonresonant process written in terms of the entanglement is a nonmonotonic function of the purities.

\onecolumngrid
\appendix

\section{Position and momentum correlation functions}

Here we present the general expression for the position and momentum correlation functions for two modes $i \in \mathcal{E}_{R}$ and $j \in \mathcal{E}_{L}$. In order to obtain the correlation function for just one mode (e.g. $\left\langle \left\{ q_{i} \left( t \right) , q_{i} \left( t \right) \right\} \right\rangle$ instead of $\left\langle \left\{ q_{i} \left( t \right) , q_{j} \left( t \right) \right\} \right\rangle$) we just need to make the replacement $\left\{ i , R \right\} \rightarrow \left\{ j , L \right\}$. The position correlator at an arbitrary environmental temperature is
\begin{equation}
	\begin{aligned}
		\left\langle \left\{ \hat{q}_{i} \left( t \right) ,  \hat{q}_{j} \left( t \right) \right\} \right\rangle & = \frac{1}{m_{i} \omega_{i}} \, \left[ 2 \, n_{R} \left( \omega_{i} \right) + 1 \right] \, \delta_{i j}\\
		& + \frac{1}{2 m} \frac{1}{\sqrt{m_{i} \omega_{i}}} \frac{1}{\sqrt{m_{j} \omega_{j}}} \Delta \omega \sqrt{I_{R} \left( \omega_{i} \right) I_{L} \left( \omega_{j} \right)} \, \left[ 2 \, n_{R} \left( \omega_{i} \right) + 1 \right] \\
		& \times \text{Im} \left[ \mathcal{J} \left( \omega_{i} , \omega_{j} , t \right) e^{- i \left( \omega_{i} - \omega_{j} \right) t} - \mathcal{J} \left( \omega_{i} , - \omega_{j} , t \right) e^{- i \left( \omega_{i} + \omega_{j} \right) t} \right] \\
		& + \frac{1}{2 m} \frac{1}{\sqrt{m_{i} \omega_{i}}} \frac{1}{\sqrt{m_{j} \omega_{j}}} \Delta \omega \sqrt{I_{R} \left( \omega_{i} \right) I_{L} \left( \omega_{j} \right)} \, \left[ 2 \, n_{L} \left( \omega_{j} \right) + 1 \right] \\
		& \times \text{Im} \left[ \mathcal{J} \left( \omega_{j} , \omega_{i} , t \right) e^{ i \left( \omega_{i} - \omega_{j} \right) t} - \mathcal{J} \left( \omega_{j} , - \omega_{i} , t \right) e^{- i \left( \omega_{i} + \omega_{j} \right) t} \right] \\
		& + \frac{1}{\sqrt{m_{i} \omega_{i}}} \frac{1}{\sqrt{m_{j} \omega_{j}}} \, \Delta \omega \sqrt{I_{R} \left( \omega_{i} \right) I_{L} \left( \omega_{j} \right)} \int_{0}^{t} d t_{1} \int_{0}^{t} d t_{2} \, \text{sin} \left[ \omega_{i} \left( t - t_{1} \right) \right] \, \text{sin} \left[ \omega_{j} \left( t - t_{2} \right) \right] \left\langle \left\{ x^{h} \left( t_{1} \right) , x^{h} \left( t_{2} \right) \right\} \right\rangle \\
		& + \frac{1}{4 m^{2}} \frac{1}{\sqrt{m_{i} \omega_{i}}} \frac{1}{\sqrt{m_{j} \omega_{j}}} \Delta \omega \sqrt{I_{R} \left( \omega_{i} \right) I_{L} \left( \omega_{j} \right)} \sum_{\alpha} \int_{0}^{\infty} d \omega \, I_{\alpha} \left( \omega \right) \, \left[ 2 \, n_{\alpha} \left( \omega \right) + 1 \right] \\
		& \times \text{Re} \left[ \mathcal{J} \left( \omega, \omega_{i} , t \right) \mathcal{J}^{\ast} \left( \omega , \omega_{j} , t \right) e^{i \left( \omega_{i} - \omega_{j} \right) t} + \mathcal{J} \left( \omega, - \omega_{i} , t \right) \mathcal{J}^{\ast} \left( \omega ,  - \omega_{j} , t \right) e^{- i \left( \omega_{i} - \omega_{j} \right) t} \right. \\
		& \left. - \mathcal{J} \left( \omega, \omega_{i} , t \right) \mathcal{J}^{\ast} \left( \omega , - \omega_{j} , t \right) e^{i \left( \omega_{i} + \omega_{j} \right) t} - \mathcal{J} \left( \omega, - \omega_{i} , t \right) \mathcal{J}^{\ast} \left( \omega , \omega_{j} , t \right) e^{- i \left( \omega_{i} + \omega_{j} \right) t} \right].
	\end{aligned}
	\label{correlatorXX}
\end{equation}
and the momentum correlator is
\begin{equation}
	\begin{aligned}
		\left\langle \left\{ p_{i} \left( t \right) ,  p_{j} \left( t \right) \right\} \right\rangle & = m_{i} \, \omega_{i} \, \left[ 2 \, n_{R} \left( \omega_{i} \right) + 1 \right] \, \delta_{i j}\\
		& + \frac{1}{2 m} \sqrt{m_{i} \omega_{i}} \sqrt{m_{j} \omega_{j}} \Delta \omega \sqrt{I_{R} \left( \omega_{i} \right) I_{L} \left( \omega_{j} \right)} \, \left[ 2 \, n_{R} \left( \omega_{i} \right) + 1 \right] \\
		& \times \text{Im} \left[ \mathcal{J} \left( \omega_{i} , \omega_{j} , t \right) e^{- i \left( \omega_{i} - \omega_{j} \right) t} + \mathcal{J} \left( \omega_{i} , - \omega_{j} , t \right) e^{- i \left( \omega_{i} + \omega_{j} \right) t} \right] \\
		& + \frac{1}{2 m} \sqrt{m_{i} \omega_{i}} \sqrt{m_{j} \omega_{j}} \Delta \omega \sqrt{I_{R} \left( \omega_{i} \right) I_{L} \left( \omega_{j} \right)} \, \left[ 2 \, n_{L} \left( \omega_{j} \right) + 1 \right] \\
		& \times \text{Im} \left[ \mathcal{J} \left( \omega_{j} , \omega_{i} , t \right) e^{ i \left( \omega_{i} - \omega_{j} \right) t} + \mathcal{J} \left( \omega_{j} , - \omega_{i} , t \right) e^{- i \left( \omega_{i} + \omega_{j} \right) t} \right] \\
		& + \sqrt{m_{i} \omega_{i}} \sqrt{m_{j} \omega_{j}} \, \Delta \omega \sqrt{I_{R} \left( \omega_{i} \right) I_{L} \left( \omega_{j} \right)} \int_{0}^{t} d t_{1} \int_{0}^{t} d t_{2} \, \text{cos} \left[ \omega_{i} \left( t - t_{1} \right) \right] \, \text{cos} \left[ \omega_{j} \left( t - t_{2} \right) \right] \left\langle \left\{ x^{h} \left( t_{1} \right) , x^{h} \left( t_{2} \right) \right\} \right\rangle \\
		& + \frac{1}{4 m^{2}} \sqrt{m_{i} \omega_{i}} \sqrt{m_{j} \omega_{j}} \Delta \omega \sqrt{I_{R} \left( \omega_{i} \right) I_{L} \left( \omega_{j} \right)} \sum_{\alpha} \int_{0}^{\infty} d \omega \, I_{\alpha} \left( \omega \right) \, \left[ 2 \, n_{\alpha} \left( \omega \right) + 1 \right] \\
		& \times \text{Re} \left[ \mathcal{J} \left( \omega, \omega_{i} , t \right) \mathcal{J}^{\ast} \left( \omega , \omega_{j} , t \right) e^{i \left( \omega_{i} - \omega_{j} \right) t} + \mathcal{J} \left( \omega, - \omega_{i} , t \right) \mathcal{J}^{\ast} \left( \omega ,  - \omega_{j} , t \right) e^{- i \left( \omega_{i} - \omega_{j} \right) t} \right. \\
		& \left. + \mathcal{J} \left( \omega, \omega_{i} , t \right) \mathcal{J}^{\ast} \left( \omega , - \omega_{j} , t \right) e^{i \left( \omega_{i} + \omega_{j} \right) t} + \mathcal{J} \left( \omega, - \omega_{i} , t \right) \mathcal{J}^{\ast} \left( \omega , \omega_{j} , t \right) e^{- i \left( \omega_{i} + \omega_{j} \right) t} \right],
	\end{aligned}
	\label{correlatorPP}
\end{equation}
where the $\mathcal{J}$ function is defined as
\begin{equation}
	\mathcal{J} \left( \omega , \omega_{i} , t \right) = \sum_{k} \int_{0}^{t} d t^{\prime} \, e^{i \left( \omega - \omega_{i} + k \omega_{d} \right) t^{\prime}} \int_{0}^{t^{\prime}} d t^{\prime \prime} \, A_{k} \left( t^{\prime \prime} \right) \, e^{- i \omega t^{\prime \prime}},
\end{equation}
which, by integrating the exponentials with a change of the order of integration, can be formally solved as
\begin{equation}
	\mathcal{J} \left( \omega , \omega_{i} , t \right) = \sum_{k} \left[ t \, \text{sinc} \left[ \left( \omega - \omega_{i} + k \omega_{d} \right) t / 2 \right] \, a_{k} \left( i \omega \right) \, e^{i \left( \omega - \omega_{i} + k \omega_{d} \right) t / 2} + F_{k} \left( \omega , \omega_{i} \right) \right],
\end{equation}
\noindent where we used the notation
\begin{equation}
	\begin{aligned}
	a_{k} \left( i \omega \right) & = \int_{0}^{t} d t^{\prime} \, A_{k} \left( t^{\prime} \right) \, e^{- i \omega t^{\prime}} \\
	F_{k} \left( \omega , \omega_{i} \right) & = \frac{a_{k} \left( i \omega \right) - a_{k} \left[ i \left( \omega_{i} - k \omega_{d} \right) \right]}{i \left( \omega - \omega_{i} + k \omega_{d} \right) }
	\end{aligned}
\end{equation}
Both $a_{k}$ and $F_{k}$ are functions of the variable $t$, but we do not write its explicit dependence in an effort to keep the notation simple. When computing correlators and quantities related to them, it is sometimes useful to use a relation which is a direct consequence of the unitary evolution of the operators $q_{i}$ and $p_{i}$:
\begin{equation}
	\begin{aligned}
		\text{Im} \left[ \mathcal{J} \left( \omega_{i} , \omega_{i} , t \right) \right] & =  - \frac{1}{4 m} \sum_{\alpha} \int_{0}^{\infty} d \omega \, I_{\alpha} \left( \omega \right) \left\lvert \mathcal{J} \left( \omega , \omega_{i} , t \right) \right\rvert^{2} + \frac{1}{4 m} \sum_{\alpha} \int_{0}^{\infty} d \omega \, I_{\alpha} \left( \omega \right) \left\lvert \mathcal{J} \left( \omega , - \omega_{i} , t \right) \right\rvert^{2} \\
		& - \frac{1}{2} m \int_{0}^{t} d t_{1} \int_{0}^{t} d t_{2} \, \text{sin} \left[ \omega_{i} \left( t_{1} - t_{2} \right) \right] \left\langle i \left[ x^{h} \left( t_{1} \right) , x^{h} \left( t_{2} \right) \right] \right\rangle.
	\end{aligned}
	\label{condicionUnitaria}
\end{equation}
The previous equality is obtained by imposing that $[ q_{i} (t) , p_{i} (t) ] = i \mathbb{1}$ for all times to the solutions of the equations of motion shown in the main text. We can identify the different parts of the covariance matrix as written in Eq. \eqref{covariance} by looking at, for example, the position correlator in Eq. \eqref{correlatorXX}. The initial thermal state is
\begin{equation}
    \boldsymbol{\nu}_{i} \otimes \boldsymbol{\nu}_{j} \rightarrow \frac{1}{2} \, \left[ 2 \, n_{R} \left( \omega_{i} \right) + 1 \right] \, \delta_{i j}.
\end{equation}
The term which involves the initial state of $\sys$ is
\begin{equation}
    \boldsymbol{\sigma}_{0} (t) \rightarrow \frac{1}{2} \, \Delta \omega \sqrt{I_{R} \left( \omega_{i} \right) I_{L} \left( \omega_{j} \right)} \int_{0}^{t} d t_{1} \int_{0}^{t} d t_{2} \, \text{sin} \left[ \omega_{i} \left( t - t_{1} \right) \right] \, \text{sin} \left[ \omega_{j} \left( t - t_{2} \right) \right] \left\langle \left\{ x^{h} \left( t_{1} \right) , x^{h} \left( t_{2} \right) \right\} \right\rangle.
\end{equation}
The other two terms, $\boldsymbol{\sigma}_{osc} (t)$ and $\boldsymbol{\sigma}_{lin} \times t$, are given by the rest of Eq. \eqref{correlatorXX}.

Using the correlators in Eq. \eqref{correlatorXX} and Eq. \eqref{correlatorPP} we can obtain $E_{i}$, which is the energy stored in the band $i \in \env_{R}$ presented in Eq. \eqref{heatCurrent}, as
\begin{equation}
    E_{i} = \frac{1}{4 m_{i}} \left\langle \left\{ p_{i} \left( t \right) ,  p_{i} \left( t \right) \right\} \right\rangle + \frac{1}{4} m_{i} \omega_{i}^{2} \left\langle \left\{ q_{i} \left( t \right) ,  q_{i} \left( t \right) \right\} \right\rangle.
\end{equation}
A direct computation using relation \eqref{condicionUnitaria} shows that
\begin{equation}
    \begin{aligned}
        E_{i} \left( t \right) & = \frac{1}{2} \omega_{i} \, \left[ 2 \, n_{R} \left( \omega_{i} \right) + 1 \right] \\
        & + \frac{1}{2} \Delta \omega \, \omega_{i} \, I_{R} \left( \omega_{i} \right) \left[ 2 \, n_{R} \left( \omega_{i} \right) + 1 \right] \int_{0}^{t} d t_{1} \int_{0}^{t} d t_{2} \, \left\langle \hat{x}^{h} \left( t_{1} \right) \hat{x}^{h} \left( t_{2} \right) \right\rangle  \, e^{- i \omega_{i} \left( t_{1} - t_{2} \right)} \\
		& - \frac{1}{2} \Delta \omega \, \omega_{i} \, I_{R} \left( \omega_{i} \right) n_{R} \left( \omega_{i} \right) \int_{0}^{t} d t_{1} \int_{0}^{t} d t_{2} \, \text{cos} \left[ \omega_{i} \left( t_{1} - t_{2} \right) \right] \left\langle \left\{ x^{h} \left( t_{1} \right) , x^{h} \left( t_{2} \right) \right\} \right\rangle \\
        & + \frac{1}{4 m^{2}} \Delta \omega \, \omega_{i} \, I_{R} \left( \omega_{i} \right) \sum_{\alpha = R,L} \int_{0}^{\infty} d \omega \, I_{\alpha} \left( \omega \right) \left[ n_{\alpha} \left( \omega \right) - n_{R} \left( \omega_{i} \right) \right] \left\lvert \mathcal{J} \left( \omega, \omega_{i} , t \right) \right\rvert^{2} \\
        & + \frac{1}{4 m^{2}} \Delta \omega \, \omega_{i} \, I_{R} \left( \omega_{i} \right) \sum_{\alpha = R,L} \int_{0}^{\infty} d \omega \, I_{\alpha} \left( \omega \right) \left[ n_{\alpha} \left( \omega \right) + n_{R} \left( \omega_{i} \right) + 1 \right] \left\lvert \mathcal{J} \left( \omega, - \omega_{i} , t \right) \right\rvert^{2}.
    \end{aligned}
    \label{energiaEi}
\end{equation}
In order to arrive at the final expression, we need the long-time behavior of the $\mathcal{J}$ functions. For this, we will use the fact that
\begin{equation}
	\lim_{t \to \infty} t \, \text{sinc} \left( \omega \, t / 2 \right) = \lim_{t \to \infty} t \, \text{sinc}^{2} \left( \omega \, t / 2 \right) = 2 \pi \, \delta \left( \omega \right),
\end{equation}
and $\lim_{t \to \infty} a_{k} ( i \omega) = \tilde{A}_{k} (i \omega)$. Thus,
\begin{equation}
    \left\lvert \mathcal{J} \left( \omega, \pm \omega_{i} , t \right) \right\rvert^{2} \to 2 \pi t \, \sum_{k} \delta ( \omega \mp \omega_{i} + k \omega_{d} ) \lvert \tilde{A}_{k} (i \omega ) \rvert^{2},
\end{equation}
where we only kept the linear terms in time. Using this, and that $\tilde{A}_{k}^{\ast} (i \omega) = \tilde{A}_{-k} (-i \omega)$ (which is a consequence of $G$ being a real valued function), we arrive at
\begin{equation}
    E_i (t) \to \left[ 1/2 + n_{R} \left( \omega_{i} \right) \right] \omega_{i} + \dot{\mathcal{Q}}_{i} \times t
\end{equation}
with $\dot{\mathcal{Q}}_{i}$ as shown in Eq. \eqref{heatCurrent} (remember that the terms involving the initial state of $\sys$ decay exponentially fast for long-times).

\section{Individual purities and $\Gamma_{\pm}$}
Here we present explicit expressions for the individual purities $\mu_{i,j}$ and for $\Gamma_{\pm}$, in the long-time regime (that is, given the covariance matrix $\boldsymbol{\sigma}_{av}$ in Eq. \eqref{covarianceAV}). As discussed above, $\mu_{i,j}$ are related to $\boldsymbol{\sigma}_{av}$ by $\mu_{i} = 1 / 2 \sqrt{\text{det} ( \boldsymbol{\alpha})}$ and $\mu_{j} = 1 / 2 \sqrt{\text{det} ( \boldsymbol{\beta} )}$. Since the purities do not depend on the relation between the frequencies $\omega_{i}$ and $\omega_{j}$, we just show
\begin{equation}
    \begin{aligned}
        2 \sqrt{\text{det} ( \boldsymbol{\alpha})} & = 1 + 2 n_{R} \left( \omega_{i} \right) \\
		& + \frac{\pi}{m^{2}} \, \Delta \omega \, I_{R} \left( \omega_{i} \right) \sum_{\alpha = R,L} \sum_{q^{\prime}} I_{\alpha} \left( \omega_{i} - q^{\prime} \omega_{d} \right) \left[ n_{\alpha} \left( \omega_{i} - q^{\prime} \omega_{d} \right) - n_{R} \left( \omega_{i} \right) \right] \left\lvert \tilde{A}_{q^{\prime}} \left[ i \left( \omega_{i} - q^{\prime} \omega_{d} \right) \right] \right\rvert^{2} \times t \\
        & + \frac{\pi}{m^{2}} \, \Delta \omega \, I_{R} \left( \omega_{i} \right) \sum_{\alpha = R,L} \sum_{q^{\prime \prime}} I_{\alpha} \left( q^{\prime \prime} \omega_{d} - \omega_{i} \right) \left[ n_{\alpha} \left( q^{\prime \prime} \omega_{d} - \omega_{i} \right) + n_{R} \left( \omega_{i} \right) + 1 \right] \left\lvert \tilde{A}_{- q^{\prime \prime}} \left[ i \left( q^{\prime \prime} \omega_{d} - \omega_{i} \right) \right] \right\rvert^{2} \times t
    \end{aligned}
    \label{muExplicit}
\end{equation}
where $q^{\prime}$ and $q^{\prime \prime}$ are integers such that $\omega_{i} - q^{\prime} \omega_{d} \geq 0$ and $q^{\prime \prime} \omega_{d} - \omega_{i} \geq 0$, respectively. In order to obtain $2 \sqrt{\text{det} ( \boldsymbol{\beta} )}$ we just need to make the replacement $\left\{ i , R \right\} \rightarrow \left\{ j , L \right\}$. On the contrary, $\Gamma_{\pm}$ depends on whether $\omega_{i} + \omega_{j} = k \omega_{d}$ or $\omega_{j} = \omega_{i} + k \omega_{d}$. Thus, for the nonresonant case ($\omega_{i} + \omega_{j} = k \omega_{d}$) we have
\begin{equation}
    \Gamma_{+} = - \frac{1}{4 m^{2}} \Delta \omega^{2} I_{R} \left( \omega_{i} \right) I_{L} \left( \omega_{j} \right) \left\lvert \left[ 2 n_{R} \left( \omega_{i} \right) + 1 \right] \tilde{A}_{-k}^{\ast} \left( i \omega_{i} \right) + \left[ 2 n_{L} \left( \omega_{j} \right) + 1 \right] \tilde{A}_{-k}^{\ast} \left( i \omega_{j} \right) \right\rvert^{2},
    \label{gmenosExplicit}
\end{equation}
and for the resonant one ($\omega_{j} = \omega_{i} + k \omega_{d}$),
\begin{equation}
    \Gamma_{-} = \frac{1}{m^{2}} \Delta \omega^{2} I_{R} \left( \omega_{i} \right) I_{L} \left( \omega_{j} \right) \left\lvert n_{R} \left( \omega_{i} \right) \tilde{A}_{k} \left( i \omega_{i} \right) - n_{L} \left( \omega_{j} \right) \tilde{A}_{-k}^{\ast} \left( i \omega_{j} \right) \right\rvert^{2}.
    \label{gmExplicit}
\end{equation}
It is clear from Eqs. \eqref{muExplicit} and \eqref{gmExplicit} that the inequality in Eq. \eqref{ineq} holds. To see this, we can write the inequality as
\begin{equation}
    \lvert \Gamma_{-} (t) \rvert \leq (1/\mu_{i}-1) (1/\mu_{j}-1).
\end{equation}
From Eq. \eqref{gmExplicit}, we know that when temperature goes to zero, the left hand side of the inequality goes to zero too. On the contrary, from the third line of Eq. \eqref{muExplicit}, the right hand side does not go to zero. In the general case, the left hand side of the inequality is of order $(\Delta \omega^{2} \gamma_{0}^{2}) n (\omega)$, while the right hand side is of order $n (\omega)^{2} + (\Delta \omega \gamma_{0}^{2} ) n (\omega)$ which is much bigger in the weak coupling limit. So, as temperature increases, the right hand side is always greater than the left hand side.

\section{Symplectic eigenvalues and measures of correlations}
Here we will briefly show how to obtain the expressions of the mutual information, quantum discord and logarithmic negativity presented above. In order to do this, first we will compute the symplectic eigenvalues of $\boldsymbol{\sigma}_{av}$. For the sake of clarity we will use the standard notation: $A = \text{det} (\boldsymbol{\alpha}) = 1 / 4 \mu_{i}^{2}$, $B = \text{det} (\boldsymbol{\beta}) = 1 / 4 \mu_{j}^{2}$, $C = \text{det} (\boldsymbol{\gamma}) = \Gamma (t) / 4$ and $D = \text{det} (\boldsymbol{\sigma}_{av})$. We remind the reader that all computations are valid in the weak coupling regime and for long times. Therefore, in many steps we will use Taylor expansions to first nontrivial order in $\lvert C \rvert$ when comparing $\lvert C \rvert \sim O(\gamma_{0}^{2})$ to $\lvert A \rvert, \lvert B \rvert \sim O (1)$ (see Eqs. \eqref{muExplicit}, \eqref{gmenosExplicit} and \eqref{gmExplicit}). The symplectic eigenvalues are
\begin{equation}
    \lambda_{1,2}^{2} = \frac{1}{2} ( \Delta \pm \sqrt{\Delta^{2} - 4 \, D} ),
\end{equation}
where $\Delta = A + B + 2 C$ \cite{adessoIlluminati,serafiniIlluminatiSiena}. In the nonresonant case we have $\Delta = A + B - 2 \lvert C \rvert$ and $D = (\sqrt{AB} - \lvert C \rvert)^{2}$ (see Eq. \eqref{standardForm}). Thus, after a Taylor expansion we obtain
\begin{equation}
    \sqrt{\Delta^{2} - 4 \, D} \simeq (A - B) - 2 \frac{\sqrt{A} - \sqrt{B}}{\sqrt{A} + \sqrt{B}} \lvert C \rvert.
\end{equation}
Therefore, after another expansion, the result is
\begin{equation}
    \begin{aligned}
        \lambda_{1} \simeq \sqrt{A} - \frac{\lvert C \rvert}{\sqrt{A} + \sqrt{B}} \\
        \lambda_{2} \simeq \sqrt{B} - \frac{\lvert C \rvert}{\sqrt{A} + \sqrt{B}}
    \end{aligned}
\end{equation}
On the other hand, in the resonant case we have $\Delta = A + B + 2 C$ and $D = (\sqrt{AB} - C)^{2}$. Then, after a Taylor expansion we get
\begin{equation}
    \sqrt{\Delta^{2} - 4 \, D} \simeq (A - B) + 2 \frac{\sqrt{A} + \sqrt{B}}{\sqrt{A} - \sqrt{B}} C.
\end{equation}
In order to arrive at the previous expression, it was used that $(\sqrt{A} - \sqrt{B})^{2} > 4C$. This in an immediate consequence of the inequality in Eq. \eqref{ineq}. The symplectic eigenvalues in this case are
\begin{equation}
    \begin{aligned}
        \lambda_{1} \simeq \sqrt{A} + \frac{C}{\sqrt{A} - \sqrt{B}} \\
        \lambda_{2} \simeq \sqrt{B} - \frac{C}{\sqrt{A} - \sqrt{B}}
    \end{aligned}
\end{equation}
Now we can proceed to compute the mutual information and quantum discord. The mutual information is defined as \cite{serafiniIlluminatiSiena}
\begin{equation}
    \mathcal{I} (\boldsymbol{\sigma}) = f (\sqrt{A}) + f (\sqrt{B}) - f(\lambda_{1}) - f(\lambda_{2}),
\end{equation}
where $f (x) = (x + 1/2) \, \text{ln} (x + 1/2) - (x - 1/2) \, \text{ln} (x - 1/2)$. Using a first order Taylor expansion in $\lvert C \rvert / (\sqrt{A} \pm \sqrt{B})$, we get
\begin{equation}
    \mathcal{I}_{\pm} (\boldsymbol{\sigma}_{av}) \simeq \pm [f^{\prime} (\sqrt{A}) \pm f^{\prime} (\sqrt{B})] \, \frac{\lvert C \rvert}{\sqrt{A} \pm \sqrt{B}}
\end{equation}
This is the expression shown in \eqref{mutualInformation}. The quantum discord is defined as \cite{adessoDatta}:
\begin{equation}
    \overleftarrow{D} ( \boldsymbol{\sigma} ) = f(\sqrt{B}) - f(\lambda_{1}) - f(\lambda_{2}) + f(\sqrt{E_{min}}).
\end{equation}
In our case, $E_{\min}$ is
\begin{equation}
    E_{\min} = \frac{2 C^{2} + (1/4-B)(A-4D) + 2\lvert C \rvert \sqrt{C^{2} + (1/4-B)(A-4D)}}{4(1/4-B)^{2}}.
\end{equation}
Replacing $D = (\sqrt{AB} - \lvert C \rvert)^{2}$ in the equation above, the expression for $E_{\min}$ can be greatly simplified. Indeed, noticing that
\begin{equation}
	\frac{2\lvert C \rvert^{2} + (1/4-B)(A-4D)}{4(1/4-B)^{2}} = \left( \sqrt{A} - \frac{\sqrt{B} \lvert C \rvert}{\lvert 1/4 - B \rvert} \right)^{2} + \left( \frac{\lvert C \rvert}{2 \lvert 1/4-B \rvert} \right)^{2}
\end{equation}
and
\begin{equation}
	\frac{2\lvert C \rvert \sqrt{C^{2} + (1/4-B)(A-4D)}}{4(1/4-B)^{2}} = \frac{\lvert C \rvert}{\lvert 1/4 - B \rvert}  \left( \sqrt{A} - \frac{\sqrt{B} \lvert C \rvert}{\lvert 1/4 - B \rvert} \right),
\end{equation}
we arrive at
\begin{equation}
    E_{min} = \left( \sqrt{A} - \frac{\lvert C \rvert}{\lvert 1/2 + \sqrt{B} \rvert} \right)^{2},
\end{equation}
where we used the fact that $1/4-B < 0$ and $\sqrt{A} > \sqrt{B} \lvert C \rvert / \lvert 1/4 - B \rvert$. Thus, using a first order Taylor expansion in $\lvert C \rvert / (\lvert 1/2 + \sqrt{B} \rvert)$, we get
\begin{equation}
    f(\sqrt{E_{min}}) \simeq f(\sqrt{A}) - f^{\prime}(\sqrt{A}) \frac{\lvert C \rvert}{\lvert 1/2 + \sqrt{B} \rvert}.
\end{equation}
Finally, the quantum discord can be written as shown in Eq. \eqref{discord}:
\begin{equation}
    \overleftarrow{D}_{\pm} ( \boldsymbol{\sigma}_{av} ) \simeq \mathcal{I}_{\pm} (\boldsymbol{\sigma}_{av}) - f^{\prime}(\sqrt{A}) \frac{\lvert C \rvert}{\lvert 1/2 + \sqrt{B} \rvert}.
\end{equation}
For the logarithmic negativity, we need the lowest symplectic eigenvalue of the transposed covariance matrix $\tilde{\boldsymbol{\sigma}}_{av}$. This is,
\begin{equation}
    \tilde{\lambda}_{2}^{2} = \frac{1}{2} ( \tilde{\Delta} - \sqrt{\tilde{\Delta}^{2} - 4 \, D} ),
\end{equation}
where $\tilde{\Delta} = A + B - 2 C$ \cite{adessoIlluminati,serafiniIlluminatiSiena}. In this case, we have
\begin{equation}
    \sqrt{\tilde{\Delta}^{2} - 4 \, D} \simeq 2 \sqrt{\lvert C \rvert}(\sqrt{A} + \sqrt{B}) \left[ 1 + \frac{(\sqrt{A} - \sqrt{B})^{2}}{8 \lvert C \rvert} \right]
\end{equation}
and therefore,
\begin{equation}
    \tilde{\lambda}_{2}^{2} \simeq \frac{1}{2} (A + B) \left[ 1 - \frac{\sqrt{A} + \sqrt{B}}{A+B} \sqrt{\lvert C \rvert} \right].
\end{equation}
The logarithmic negativity as shown in Eq. \eqref{entanglement} is obtained after computing $\text{max} \{ 0 , - \text{ln} (2 \tilde{\lambda}_{-}) \}$ and using the fact that $\text{ln} (1 + x) \simeq x$ for small $x$.

\twocolumngrid

\end{document}